\documentclass[aps,prl,preprint,showpacs,superscriptaddress]{revtex4}
\usepackage[dvips]{color}
\definecolor{red}{rgb}{1,0,1}

\def \beq{\begin{equation}}         \def \eeq{\end{equation}}
\def \beqa{\begin{eqnarray}}        \def \eeqa{\end{eqnarray}}
\def \bea{\begin{array}}        \def \eea{\end{array}}

\def\bio#1#2#3{{Biophys. J. }{\bf #1}, #2 (#3)}
\def\biom#1#2#3{{Biomacromolecules }{\bf #1}, #2 (#3)}
\def\biop#1#2#3{{Biopolymers }{\bf #1}, #2 (#3)}
\def\mol#1#2#3{{Macromolecules }{\bf #1}, #2 (#3)}
\def\epje#1#2#3{{Eur. Phys. J. E }{\bf #1}, #2 (#3)}
\def\sm#1#2#3{{Soft Matter }{\bf #1}, #2 (#3)}
\def\pnas#1#2#3{{Proc. Natl. Acad. Sci. USA }{\bf #1}, #2 (#3)}

\def\prl#1#2#3{{Phys. Rev. Lett. }{\bf #1}, #2 (#3)}
\def\nat#1#2#3{{Nature }{\bf #1}, #2 (#3)}
\def\slb#1#2#3{{Phil. Trans. R. Soc. Lond. B }{\bf #1}, #2 (#3)}

\def\jeb#1#2#3{{J. exp. Biol. }{\bf #1}, #2 (#3)}
\def\jmb#1#2#3{{J. Mol. Biol. }{\bf #1}, #2 (#3)}
\def\sci#1#2#3{{Science }{\bf #1}, #2 (#3)}

\def\jps#1#2#3{{J. Polymer Sciences: Polymer Symposium }{\bf #1}, #2 (#3)}
\def\trj#1#2#3{{Textile Res. J. }{\bf #1}, #2 (#3)}
\def\ja#1#2#3{{J. Arachnol. }{\bf #1}, #2 (#3)}

\usepackage{amsmath}
\usepackage{epsfig}

\begin{document}

%Title of paper
\title{Elasticity of Spider Dragline Silks Viewed as Nematics: Yielding Induced by Isotropic-Nematic Phase Transition
}
\author{Lin-ying Cui}
\email[Email address:]{cly05@mails.tsinghua.edu.cn}
\affiliation{Department of physics, Tsinghua University, Beijing,
100084, China}
\author{Fei Liu}
\affiliation{Center for Advanced Study, Tsinghua University,
Beijing, 100084, China}
\author{Zhong-can Ou-Yang}
\affiliation{Center for Advanced Study, Tsinghua University,
Beijing, 100084, China}\affiliation{Institute of Theoretical
Physics, The Chinese Academy of Sciences, P.O.Box 2735 Beijing
100190, China}

\date{\today}

\begin{abstract}
{Spider dragline silk shows well-known outstanding mechanical
properties. However, its sigmoidal shape of the measured
stress-strain curves ({\it i.e.} the yield) can not be described
 by classical polymer theories and recent hierarchical
chain model. To solve the long lasting problem, we generalized the
Maier-Saupe theory of nematics to construct an elastic model for the
polypeptide chain network of the dragline silk. The comprehensive
agreement between theory and experiments on the stress-strain curve
strongly indicates the dragline silks to belong to liquid crystal
elastomers. Especially, the remarkable yielding elasticity of the
silk is understood for the first time as the force-induced
isotropic-nematic phase transition of the chain network. Our theory
also predicts a drop of the stress in supercontracted dragline silk,
an early found effect of humidity on the mechanical property in many
silks. }
\end{abstract}
 \maketitle

Spider dragline silks (SDSs), the main structural web silk regarded
as the ``spider's lifeline", exhibit fascinating mechanical
properties, such as a tactful combination of high tensile strength
and high extensibility~\cite{gosline}, thus showing a remarkably
sigmoidal shape of the measured stress-strain
curves~\cite{Vollrath2}. Several experimental studies have been
carried out to determine the supra-molecular structure organization
of the SDS~\cite{Anja,Hayashi,Oroudjev,Simmons,Krasnov} and tried to
produce mimic silks with similar properties~\cite{Li}. It is now
widely accepted that SDSs are semicrystalline polymers with
$\beta$-sheet nanocrystals embedded in amorphous region, which is a
polypeptide chain network~\cite{Termonia,Vollrath1,Lefevre}; see
Fig. 1(a). However, the deformation mechanism, which is essential
for understanding the SDS's extraodinary mechanical properties and
mimicking the silk, is still in intense
debate~\cite{Anja,Porter,Vollrath1,Vehoff,Papa}.

On the theoretical side, to understand the exceptional mechanical
properties of SDS is of longstanding interest, and many models have
been proposed~\cite{Termonia,Porter,Vollrath1,Krasnov} and some
insights attained~\cite{Termonia,Porter,Vollrath1,Krasnov}. For
example, the model by Termonia~\cite{Termonia} treated the SDS as a
hydrogen-bonded amorphous region embedded with stiff crystals as
cross-links. In the interfacial region, an extremely high modulus is
required to get the dragline's overall behavior on deformation.
While, in the model of Porter and Vallrath~\cite{Porter,Vollrath1},
parameters linking to chemical compositions and morphological order
were used to interpret thermo-mechanical properties. But some
parameters such as ordered/disordered fractions are difficult to be
obtained from experiments. A recent model~\cite{Krasnov} connecting
deformations on macroscopic and molecular length scales still did
not consider the change of the orientation of nanocomposites during
deformation. Especially, as pointed out by Vehoff {\it at al.}
recently~\cite{Vehoff}, basic polymer theories such as the freely
jointed chain, the freely rotating chain and the worm-like chain, as
well as a hierarchical chain model of spider capture
silk~\cite{haijun} can not reproduce the sigmoidal shape or even the
steep initial regime of the spider dragline silk [ Fig.
2(a)]~\cite{Vehoff}. In one word, a unified description for SDS as a
model biomaterial still seems to be lacking.

Quite a few
works~\cite{Vollrath2,Knight,Anja,Simmons,Li,Kerkam,willcox} have
pointed out that spider silk is liquid crystalline material and
liquid crystal (LC) phase plays a vital role in both its spinning
process and mechanical properties. In the spinning process, the
liquid crystalline `spinning dope' helps spider to control the
folding and crystallization of the main protein constituents at
benign condition (close to ambient temperatures and pressures using
water as solvent)~\cite{Vollrath2,Kerkam,willcox}. The liquid
crystalline phase also plays an important role in the solid silk's
properties~\cite{Knight}. For instance, several works have found out
that the orientation of nanocomposites can affect SDS's mechanical
properties significantly~\cite{Anja,Papa,Simmons}. A recent
experiment also suggested the existence of conformational transition
and the liquid crystalline state of regenerated silk fibroin in
water~\cite{Li}. Therefore, to present an analytically tractable LC
model of the SDS that can catch the main physical factors is a
current challenge to theorists. In this work, we generalized the
Maier-Saupe theory~\cite{Maier} of nematic LC to construct an
elastic model for the polypeptide chain network of the SDS. We show
that on deformation the SDS undergoes significant changes with
orientation of the chain network increased and the dimension of the
silk along force direction elongated. The comprehensive agreement
between theory and experiments on the stress-strain curve strongly
indicates the SDSs to belong to LC elastomers, described as a new
class of matter recently~\cite{warner}. Especially, the remarkable
yielding elasticity of the SDS is understood for the first time as
the force-induced isotropic-nematic phase transition of the chain
network and the self-consistently obtained yield point agrees with
experimental data well. The present theory also predicts a drop of
the stress in supercontracted SDS, an early found effect of humidity
on the mechanical properties in many
silks~\cite{Bell,Vehoff,Vollrath1}.

We take the polypeptide chain network in the amorphous region of the
SDS as a molecular LC field with each chain section corresponding to
a mesogenic molecule; see Fig. 1. Because the SDS's high
extensibility results primarily from the disordered
region~\cite{Krasnov,Oroudjev,Hayashi}, and many experiments showed
that the deformation of the crystals is at least a factor of 10
smaller than that of the bulk~\cite{Anja} and that the orientation
of the $\beta$-sheets is almost unchanged (usually very high) under
stress ~\cite{Philip,Papa}, we can neglect the deformations and
rotations of the $\beta$-sheet crystals in current work.

Following the LC continuum theory in the absence of forces, the
potential of a mesogenic molecule takes the Maier-Saupe interaction
form~\cite{Maier}
\begin{eqnarray}
V(\cos\theta)=-aS(\frac{3}{2}\cos^2\theta-\frac{1}{2}),
\end{eqnarray}
where $\theta$ is the angle between the long axis of the molecule
and the silk axis (the z-axis), which is also the direction of
$\hat{\bf n}$ [ Fig. 1 (b)], $a$ is the strength of the mean field,
and $S$ is the orientation order parameter of the LC, defined as the
average of second Legendre polynomial~\cite{gennes}
\begin{eqnarray}
S=\left\langle\frac{3}{2}\cos^2\theta-\frac{1}{2}\right\rangle.
\end{eqnarray}
We notice that the Maier-Saupe potential has been used by Pincus and
de Gennes in investigating LC phase transition in a polypeptide
system~\cite{Pincus}. When a uniform force field $\bf f$ along
z-axis is applied, the potential of a molecule is written as
\begin{eqnarray}
U(\cos\theta)=V-f l \cos\theta,
\end{eqnarray}
where $l$ denotes the length of the mesogenic molecule.

From the definition of the order parameter $S$, we get a
self-consistency equation
\begin{eqnarray}
\begin{array}{lll}
S&=&\int_{-1}^{1}(\frac{3}{2}\cos^2\theta-\frac{1}{2})\exp(\frac{3aS}{2{\rm
k_BT}}\cos^2\theta+\alpha \cos\theta)d\cos\theta\\
&&\left/\int_{-1}^{1}\exp(\frac{3aS}{2{\rm k_BT}}\cos^2\theta+\alpha
\cos\theta)d\cos\theta\right.,
\end{array}
\end{eqnarray}
with $\alpha=f l/{\rm k_B}T$. The solution of the above equation may
not be unique, in order to obtain physically sound solution we still
need the requirement of minimization of the free energy given by
\begin{eqnarray}
F_{MS}=-{\rm k_B}T \ln Z+\frac{1}{2}a S^2,
\end{eqnarray}
where $Z$ is the partition function
$Z=\int_{-1}^{1}e^{-U(\cos\theta)/{\rm k_B}T}d\cos\theta$, and the
second term at the right-hand side corrects for the double counting
arising from the mean field method~\cite{warner}.

We calculate the orientation function $S$ numerically at temperature
$T^*=T/T_{\rm ni}$ and force $f$ and show the results in Fig. 2(b).
Here $T_{\rm ni}=a/(4.541\rm k_B)$ is the isotropic-nematic
transition temperature in the absence of forces~\cite{Maier}. We see
that, at temperatures below $T_{\rm ni}$ the molecules have
spontaneous nematic order, and the force does not induce further
order significantly. While for the molecules initially in
paranematic states, the applied force field will induce a
first-order phase transition, which means $S$ jumps discontinuously
to a higher value at a certain critical force $f_{\rm C}(T^*)$. At
even higher temperatures, nematic field is weaker and the effect of
the force is less dramatic. Interestingly, the $\alpha-S$ curves at
different temperatures are qualitatively similar to the
stress-orientation curves given by a much more complex nematic
elastomer theory~\cite{warner2} (Fig. 5 in Ref. [25]).

To compare with the mechanical experiments of the SDS, we give the
expressions for stress and strain in our theoretical framework.
Apparently, the stress $\sigma$ of a bulk is $\sigma\equiv F/A=N f$:
$F$ is the force on the surface of the bulk, $A$ is the area of the
surface, and $N$ is the number of molecules per area. The strain
$\varepsilon$ of the bulk is defined as
$\varepsilon=[L(f)-L_0]/L_0$, where $L(f)$ is the length of the bulk
along z-axis when the force field $\bf f$ is applied and we can take
it as $L(f)=\langle l|\cos\theta|\rangle$, and $L_0=L(f=0)=l/2$.
Then the strain $\varepsilon$ is $\varepsilon=2\langle|\cos
\theta|\rangle-1$. We show $\varepsilon$ versus $\sigma$ at
different temperatures in Fig. 2(c). We see that at temperatures
below $T_{\rm ni}$, the strain grows smoothly with the stress. While
for temperatures just above $T_{\rm ni}$, the strain grows with the
stress in almost a linear way under small forces, and then a jump in
the strain occurs at the critical force $f_{\rm C}(T^*)$, after
which the strain increases smoothly with the stress again. At even
higher temperatures, the jump is replaced by a smooth increase in
strain, but there is a plateau in a certain range of force.

In our model, the reduced temperature $T^*$ is an essential
parameter, and we need to choose a proper value for it in order to
predict the stress-strain curve of the SDS. Experiments showed that
the solution from which the SDS was drawn was in liquid crystalline
state at ambient temperature and
pressure~\cite{Vollrath2,Kerkam,willcox,Li}, while the orientation
in the amorphous region of solid silk was very low~\cite{Papa}.
Thus, we assign the isotropic-nematic transition temperature $T_{\rm
ni}$ of the cross-linked chain network slightly lower than the room
temperature $T_r$. Namely, the SDS is in paranematic state at
ambient temperature and $T^*$ is just above 1. The
$\sigma-\varepsilon$ curves in the paranematic states in our model
indeed exhibit main features of the stress-strain relation of the
SDS: there is a linear increase in stress with strain at small
values, and then at a certain strain and afterwards, the material
becomes softer with lower Young's modulus~\cite{Knight,Du}. We then
reveal the beginning of the isotropic-nematic phase transition as
the yield point. We see that the curves with $T^*=1.01$ and 1.02
agree well with the measured curve in the beginning linear region
and the yield point. But because the actual deformation process of
the silk is more complicated and additional factors may be involved,
such as the viscoelasticity, defects and poly-domain
effect~\cite{warner},the curve with $T^*=1.1$ agrees better with the
overall stress-strain measurement topologically. We calculate the
yield point by choosing curves with $T^*=1.01$ and 1.02. We get the
yield strain $\varepsilon_y\approx0.04$, the yield stress
$\sigma_y=\alpha N {\rm k_B}T/l\approx8.4 {\rm MPa}$, and the
Young's modulus at the linear region
$E\equiv\sigma_y/\varepsilon_y\approx210 {\rm MPa}$, given
$\alpha\sim0.2$, $N/l\sim10 {\rm nm}^{-3}$, and ${\rm k_B}T_r\sim4.1
{\rm pN nm}$. These results agree with experimental
data~\cite{Du,Vollrath1} [Fig. 2(d)] satisfactorily.

We notice that our results agree much better with the mechanical
properties of the silks with low spinning speed. That is because the
spinning speed can induce a low orientation in the amorphous region
which makes the silk more stiff. Since this additional order in the
amorphous region is not taken into account in the current work, the
silks in our model are generally a bit softer than the silks with
high spinning speed. Another thing needs pointing out is that we
predict there is a phase transition at the yield point which is
supported by a few experiments. For instance, in the polarized FTIR
spectroscopy experiment by Papadopoulos {\it et al.}, the
orientation of some components in the amorphous region increased by
0.3 when the strain reached 24\%. Besides, we would like to discuss
about the isotropic-nematic transition temperature $T_{\rm ni}$ of
the cross-linked chain network. If we choose $T^*=1.1$ in our
calculation, the transition temperature $T_{\rm ni}\approx270{\rm
K}$, which is reasonable.

In addition to describing the stress-strain relation of the SDS, our
simple theory can also qualitatively account for the drop of the
stress in the wet SDS, {\it i.e.} the supercontracted SDS. We take
$L_0$ and $R_0$ as the initial length and radius of the silk, and
$L$ and $R$ as those under stress. Under the assumption of volume
conservation we have $\pi R_0^{2}L_0=\pi R^{2}L$, so
$R/R_0=\sqrt{1/(1+\varepsilon)}$. The free energy of the bulk can be
written as
\begin{eqnarray}
\begin{array}{lll}
F&=&U-f_{ext}(L-L_0)+2\pi R L\gamma\\
&=&U-f_{ext}(L-L_0)+2\pi R_0L_0\gamma \sqrt{1+\varepsilon},
\end{array}
\end{eqnarray}
where $U$ is the internal energy of the bulk, $f_{ext}$ is the
external force on the bulk and $\gamma$ is the surface energy
coefficient. Minimizing $F$ with respect to $\varepsilon$, we get
\begin{eqnarray}
\sigma=\frac{f}{\pi R_0^{2}}=\frac{1}{\pi R_0^{2}L_0}\frac{\partial
U}{\partial \varepsilon }+\frac{\gamma}{R_0\sqrt{1+\varepsilon}}.
\end{eqnarray}
When the silk is immersed in water, the surface energy coefficient
$\gamma$ increases, so with the same stress $\sigma$ we get a bigger
strain. Thus our theory can predict the softening of supercontracted
silk, an effect observed in many
experiments~\cite{Bell,Vehoff,Vollrath1,work2,work3}.

In conclusion, we investigate the mechanical properties of the SDS
from a point of view of the LC continuum theory. We found out that
the deformation process is a force-induced isotropic-nematic phase
transition process. Remarkably, such a simple model with Maier-Saupe
theory is able to reproduce the stress-strain curve of the SDS, get
the yield point, and qualitatively interpret the drop of the stress
in the supercontracted silk.

This work is supported by the National Innovation Research Project
for Undergraduates.

\newpage
\begin{figure}[htpb]
\begin{center}
\includegraphics[width=1\columnwidth]{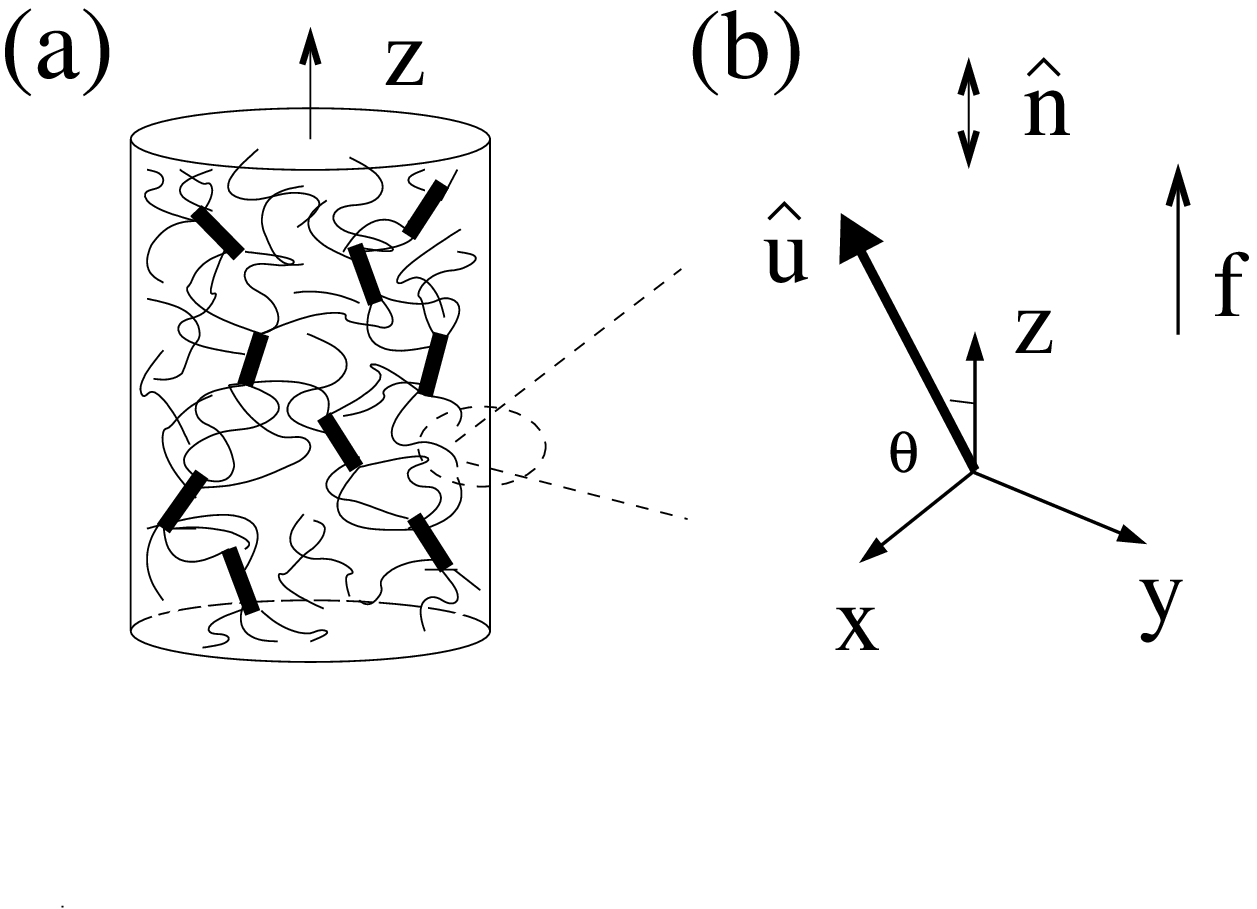}
\caption{ (a). A schematic diagram of the structure of the dragline
silk. The bold lines represent the $\beta$-sheet crystals, and the
thin lines represent the polypeptide chains in the amorphous region.
The z-axis is along the silk axis. (b). The coordinate system of the
nematics. $\hat{\bf n}$ is the director of the nematics, $\hat{\bf
u}$ is the director of the mesogenic molecule, and $\theta$ is the
angle between the long axis of the molecule and the silk axis $z$.
}\label{figure1}
\end{center}
\end{figure}

\begin{figure}[htpb]
\begin{center}
\includegraphics[width=1\columnwidth]{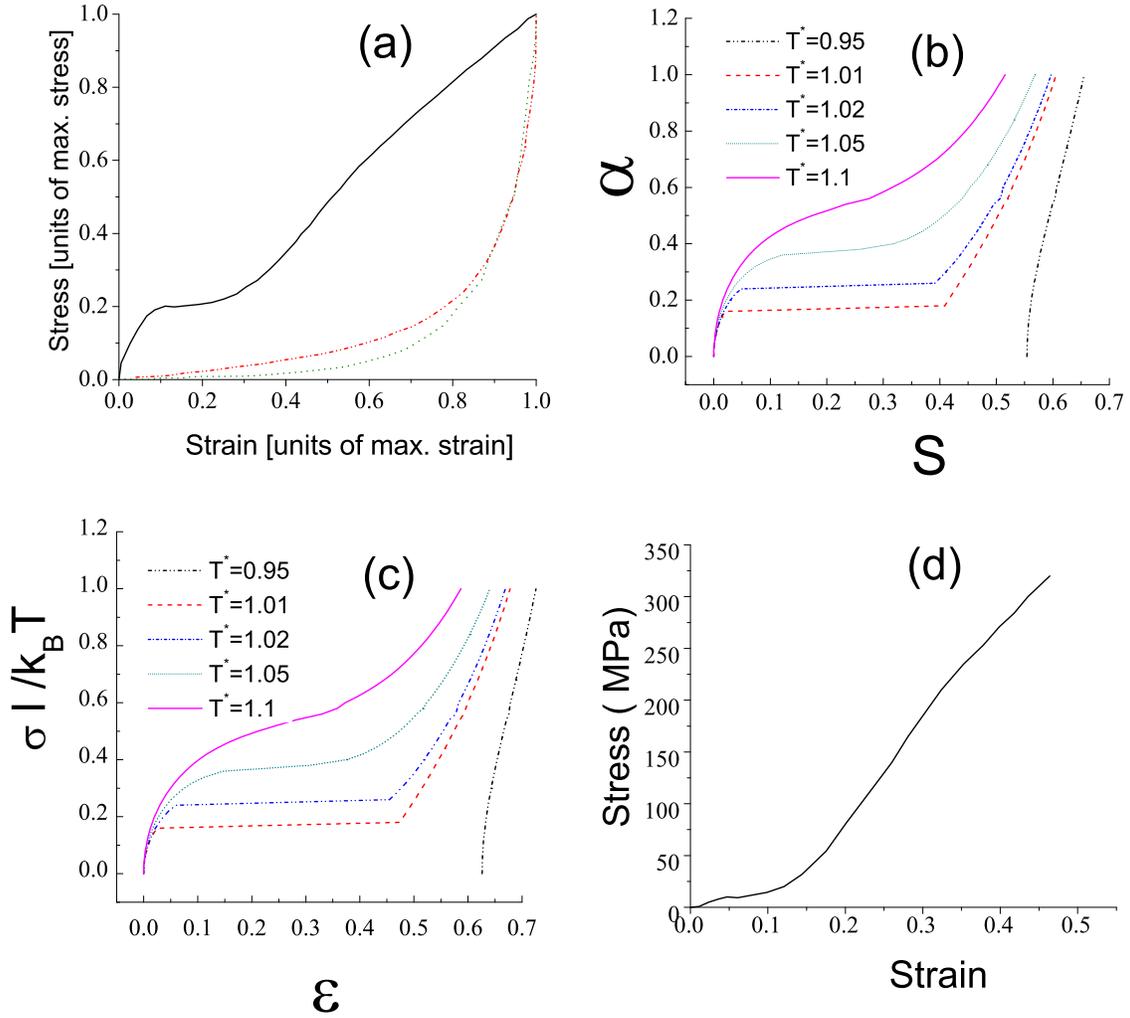}
\caption{(Color online.) (a) Comparison of a typical measured
dragline silk's stress-strain curve (black solid line) with
theoretical curves evaluated by the freely jointed chain (red dash
dot line) and the hierarchical chain model (olive dash line) [After
T. Vehoff {\it et al.}]. (b) The orientation order parameter $S$ as
a function of $\alpha$($=f l/{\rm k_B}T$). (c) The stress-strain
curves at different temperatures $T^*(=T/T_{\rm ni})$. (d) The
stress-strain curve of the SDS spinned with the speed of $1 {\rm
mms}^{-1}$ [After N. Du {\it et al.}]. }\label{figure2}
\end{center}
\end{figure}
\end{document}